\documentclass[pra,bibnotes,twocolumn]{revtex4}
\usepackage{graphicx}
\begin{document}
\draft
\def\ds{\displaystyle}
\title{Eigenstate clustering around exceptional points }
\author{ Cem Yuce  }
\address{ Department of Physics, Eskisehir Technical University, Eskisehir, Turkey }
\email{cyuce@eskisehir.edu.tr}
\date{\today}
\begin{abstract}
We propose an idea of eigenstate clustering in non-Hermitian systems. We show that non-orthogonal eigenstates can be clustered around exceptional points and illustrate our idea on some models. We discuss that exponential localization of eigenstates at edges due to the non-Hermitian skin effect is a typical example of eigenstate clustering. We numerically see that clustering of localized or extended eigenstates are possible in systems with both open and closed boundaries. We show that gain and loss can enhance eigenstate clustering. We use fidelities and the standard k-means clustering algorithm for a systematic study of clustered eigenstates. 
\end{abstract}
\maketitle

\section{Introduction}

Recently, the non-Hermitian skin effect (NHSE) has been introduced \cite{nhs1} in the study of topological phase of non-Hermitian systems \cite{reva1,reva2,reva3,reva4,reva420}. It was shown that not only topological edge states but also bulk states are exponentially localized around the boundaries in a nonreciprocal lattice with open edges. This extensively large density of eigenstates at edges implies that the standard bulk-boundary correspondence based on Bloch band topological invariants fails \cite{nhs3,nhs2,nhs4,nhs5,nhs6,nhs7,nhs8,nhs9,nhs10,nhs11,nhs12,nhs13,nhs14,nhs15,nhs16,nhs17,refek2,refek3,refek4}.  The NHSE has recently been observed in some experiments \cite{nhs18,nhs19,nhs20,nhs21}. \\
One of the unique character of non-Hermitian systems is the appearance of exceptional points (EPs). An EP is a topological singular point in the parameter space of a non-Hermitian Hamiltonian and occurs when at least two eigenvalues and the corresponding eigenvectors of the Hamiltonian coalesce \cite{kato}. An EP determines a phase transition from a real spectrum to a complex spectrum. A given non-Hermitian Hamiltonian at an EP can be brought to a matrix containing at least one Jordan block via a similarity transformation. It was recently shown that an EP appears if the non-Hermitian Hamiltonian is nilpotent \cite{CYHR}.\\
In this Letter, we propose an idea of eigenstate clustering in non-Hermitian systems. In Hermitian systems, all eigenstates are linearly dependent from each other and hence no eigenstate clustering occurs. However, eigenstates are generally non-orthogonal in non-Hermitian systems and in some certain cases all eigenstates are densely packed together and form a cluster. Therefore classifications of eigenstates are necessary in non-Hermitian systems. Here, we develop an idea of classification of non-orthogonal eigenstates into groups, where eigenstates in the same group are quite similar to each other. We illustrate our idea and specifically use k-means clustering algorithm to study clustering systematically. We claim that exponential localization of all eigenstates due to the non-Hermitian skin effect is just a typical example of eigenstate clustering. We discuss that eigenstate clustering is not restricted to localized eigenstates and show that extended eigenstates can also be densely packed in non-Hermitian systems. We further show that eigenstate clustering can also be seen in systems with closed boundary. We find that EPs play a major role for eigenstate clustering. 

 \section{Eigenstates Clustering}

Eigenstates are generally nonorthogonal in non-Hermitian systems with some exceptions such as anti-Hermitian Hamiltonians, $\ds{\mathcal{H}=-\mathcal{H}^{\dagger}}$, whose eigenstates with purely imaginary eigenvalues are all orthogonal to each other \cite{cvd1}. As opposed the orthogonal eigenstates, nonorthogonal eigenstates are not linearly dependent. In some extreme cases, non-orthogonal states become even so close to each other that one can hardly distinguish them. As an example, consider a highly nonreciprocal 1D lattice with open edges where the contrast between the hopping amplitudes in the forward and backward directions are quite large \cite{nhs1,nhs3}. In this case, the densities of all eigenstates have almost the same form and are densely populated at either edge due to the non-Hermitian skin effect. Here, our aim is to find the general condition of such closeness of eigenstates for a given non-Hermitian Hamiltonian. Suppose that non-orthogonal eigenfunctions $\ds{\psi_n  }$ for a non-Hermitian Hamiltonian are only slightly different from each other. Therefore we expand the wave function for the system order by order around a base function as follows
\begin{equation}\label{f9wl01m}
 \psi_n=\psi_B+\epsilon~ \psi^{(1)}_n +\epsilon^2 ~\psi^{(2)}_n+...
\end{equation}
where $\ds{\psi_B  }$ is the base state, $\epsilon$ is a very small parameter and $\ds{ \psi^{(1)}_n}$ and $\ds{ \psi^{(2)}_n}$ are the first and second order contributions. An important distinguishing difference between our approach and the standard perturbative treatment in quantum theory is that the zeroth order eigenstate $\ds{\psi_B  }$ is independent of $n$ in our case. One can also expand the corresponding eigenvalues in the similar way.\\
A question arises. What is the condition satisfied by non-Hermitian Hamiltonians whose eigenstates can be expressed by the above expansion (\ref{f9wl01m})? To answer this question, we start with the idea of exceptional points. Consider an $N$-level non-Hermitian Hamiltonian.  An N-th order exceptional point is a point singularity in the parameter space of the system at which all eigenstates and their eigenvalues coalesce. At such a point, the above expansion takes a simple form:  $\ds{ \psi_n=\psi_B}$. This implies that the base state is in fact the exceptional state. Consider now that the Hamiltonian is perturbed around the second order exceptional point: $\ds{  \mathcal{H}= \mathcal{H}_{EP} +\epsilon^2~ \mathcal{H}_1 }$, where $\ds{\psi_B  }$ is the only eigenstate of $\ds{ \mathcal{H}_{EP} }$, which has a Jordan block form at the exceptional point and $\ds{\epsilon^2~ \mathcal{H}_{1} }$ is the perturbative term. In this case no more coalescence of eigenstates occurs and one can analytically find the shift in the form of the eigenstate perturbatively using (\ref{f9wl01m}). We stress that the Hamiltonian goes with $\ds{\epsilon^2}$. Let us now visualize the probability densities of the eigenstates $\ds{|\psi_n|^2}$. Since $\ds{ \psi^{(1)}_n}$ and $\ds{ \psi^{(2)}_n}$ contribute perturbatively, all eigenstates are densely packed. In other words, for very small values of $\ds{\epsilon}$ (when the system is around exceptional point), the probability densities of all eigenstates are so accumulated in some regions that they can be hardly distinguishable. This eigenstate clustering is unique to non-Hermitian systems. Based on our discussion, the exceptional state $\ds{\psi_B  }$ can also be called the clustering function since all eigenstates are clustered around it (in its neighbourhoods). A typical example of eigenstate clustering is the non-reciprocal lattice with open edges, where all eigenstates are tightly localized around either edge due to the non-Hermitian skin effect. We stress that eigenstate clustering is not restricted to this example and can also occur in various systems. For example, eigenstates clustering is possible in some closed nonreciprocal lattices (with no open edges) and even reciprocal lattices with gain and loss. Clustered eigenstates needn't to be localized and they can be extended all over the lattice. Furthermore, we note that clustered eigenstates can be localized not only at around edges but also at around any point in the system. Below, we will discuss them in detail. To this end, we note that the number of clustered eigenstates rapidly decreases as we go away from the exceptional point.

\subsection{Clustering algorithm}

Clustering as a statistical data analysis is extensively used in Machine Learning. It is an unsupervised learning method since we don't need to label data points during learning. One can benefit a clustering algorithm to classify each unlabeled data into a specific group on the basis of similarity and dissimilarity between them. Data points in the same group are expected to have similar properties. On the other hand, data points in different groups should have different properties. Above, we have qualitatively discussed the concept of eigenstates clustering. The next step is to cluster eigenstates systematically according to their similarities. To do this, we need a parameter. Here we use the fidelity as a parameter to measure closeness of non-orthogonal eigenstates to each other. One can numerically calculate fidelities among eigenstates and then use a clustering algorithm to  study eigenstate clustering as a function of some parameters of a given Hamiltonian. Let us start with the definition of the fidelity
\begin{equation}\label{vn6dsvn22}
F_{nm}= \frac{|<\psi_n|\psi_m>|^2}{  <\psi_n|\psi_n> <\psi_m|\psi_m>  }
\end{equation}
where  $\ds{|\psi_n>}$ and $\ds{|\psi_m>}$ are two distinct eigenstates (In quantum information, fidelities are generally calculated for mixed states. Here we restrict ourself to eigenstates). Zero fidelity between two eigenstates means that they are orthogonal to each other. If it is close to $1$, the corresponding eigenstates are hardly distinguishable. Fidelities can not be greater than $1$.\\
In Hermitian systems, fidelities between any two distinct eigenstates are always zero. However, in non-Hermitian systems, they can have values in the interval $\ds{   [  0,1  )   } $. As a limiting case, consider an N-th order exceptional point in an N-level system. All eigenstates coalesce at such an exceptional point and hence we can not define the fidelity between two distinct states. Around the exceptional point, the forms of the eigenstates become slightly different from each other since the fidelities are piled up around $1$, which is the signal of eigenstate clustering. As we will show below as an example, the fidelities between any two distinct eigenstates are close to $1$ when the non-Hermitian skin effect occurs. \\
Having discussed the fidelity parameter, we are in a position for the systematic study of grouping the eigenstates in such a way that the eigenstates in the same group are more similar to each other than to those in other groups. In other words, the functional forms of any two eigenstates in two different groups should be different from each other as much as possible to have more meaningful grouping. To do this, we categorize non-orthogonal eigenstates into classes. There are many clustering algorithms available in the literature and many of them have already been used extensively in machine learning \cite{mlbook}. For our system, one can use various methods such as principal component analysis, the k-nearest neighbors algorithm and the k-means clustering algorithm to divide all the eigenstates into separate groups, called clusters. Here we use the k-means clustering algorithm, where $k$ is the number of classes  \cite{mlbook}. It is a relatively simple, convergence-guaranteed, computationally fast and efficient technique compared to other clustering algorithms. But choosing the $k$ value manually is the disadvantage of this technique. Let us consider an N-level non-Hermitian Hamiltonians. Suppose that there exists a set eigenstates $\ds{ \{  | \alpha>\}= \{ \psi_1,\psi_2,...,\psi_{N^{\prime}}   \}   }$, where the fidelities among them are either exactly or almost equal to zero and $\ds{N^{\prime}}$ is the total number of such eigenstates which can be found. In this way, we are able to construct $\ds{N^{\prime}}$-dimensional data space. This can be achieved by calculating the fidelities for the rest of the eigenstates with respect to $\ds{ \{ | \alpha>\}}$. After producing a set of data, we finally apply the standard k-means clustering algorithm to group the eigenstates. Note that clustering becomes trivial if $\ds{N^{\prime} =1}$. One can also make an extension of our approach to $N$-level Hermitian systems, where $\ds{N^{\prime} =N}$ (the space is in fact Hilbert space). In this case, the  number of classes is equal to the total number of eigenstates according to k-means clustering algorithm $\ds{k=N}$. This means that no eigenstate clustering occurs in Hermitian systems. Below, we illustrate our idea on some examples. 
\begin{figure}[t]\label{2678ik0}
\includegraphics[width=4.25cm]{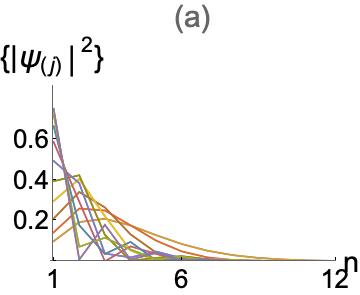}
\includegraphics[width=4.25cm]{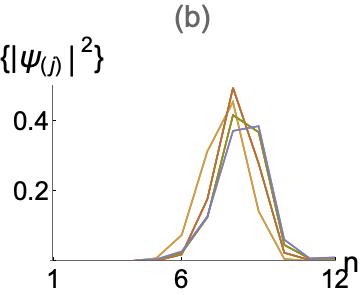}
\includegraphics[width=4.25cm]{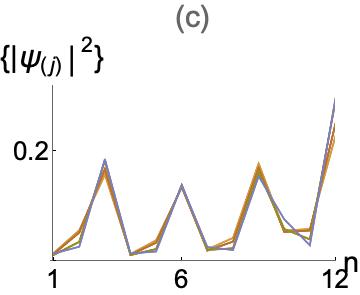}
\includegraphics[width=4.25cm]{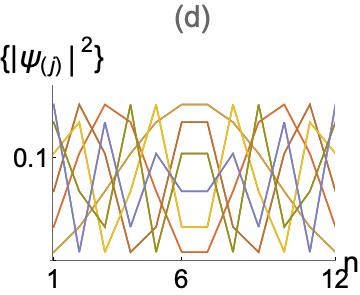}
\caption{ The plots show the set of densities of all eigenstates $\ds{ \{|\psi_{(j)} |^2\}   }$, where $j=1,2,...,N$ and $\ds{ \psi_{(1)}    }$ is the lowest lying state and $\ds{ \psi_{(N) } }$ is the state with the highest eigenvalue. We assume that there are $N=12$ lattice sites. Each color represents a distinct eigenstate. The clustered eigenstates can be seen in (a,b,c) and no clustering is possible for the Hermitian one in (d). The parameters are $t_n=0.1$ and $\gamma_n=0$ for all plots, but the backward hopping amplitudes are different for each plots: $t^{\prime}_n=0.05$ for (a), $t^{\prime}_n=0.1+\sin^2(n/3)$ (b) and $t^{\prime}_n=0.1+\sin^2(n)$ (c). The system in (a) has open edges with $t_0=t_N=0$ and hence all of the eigenstates are localized around the left edge. This is due to the non-Hermitian skin effect. However, the system boundaries in (b,c) are closed with $t_0=t_N=0.1$. The clustered eigenstates in (b) are localized but not at the edge. However, the clustered eigenstates in (c) are extended and tightly packed in such a way that it is hard to distinguish them. }
\end{figure}

\subsubsection{Examples}

Consider a generic one dimensional lattice with $N$ sites. The system has gain/loss impurities and site dependent nearest neighbouring hopping amplitudes $\ds{t_n}$ and $\ds{t^{\prime}_n}$ in the forward and backward directions, respectively \cite{refek1}.
\begin{equation}\label{hamfhksop}
\mathcal{H}\psi_n= t_n \psi_{n+1}+t^{\prime}_{n-1} \psi_{n-1}+i\gamma_n \psi_n
\end{equation}
where $\gamma_n$ is site dependent gain/loss strength. The system is Hermitian if $\ds{t_n=t^{\prime}_n}$ and $\gamma_n=0$ for all $\ds{n}$. One can see that an N-th order exceptional point occurs at either $\ds{t_n=0   }$ or $\ds{t^{\prime}_n=0    }$ when $\ds{\gamma_n=0}$. Note that there may exist other combinations of the parameters at which such an exceptional point appear. For example, one can get an exceptional point in the reciprocal lattice $\ds{t_n=t^{\prime }_n  }$ with $N=2$ by varying the gain/loss strength $\gamma_n$. \\
We perform some numerical calculations. In Fig. 1, we consider four special cases when there are $N=12$ lattice sites and plot the densities of all eigenstates for each cases. One can see that the densities are densely populated in the first three plots while no such clustering exists for the last one. In fact, the last one is for the Hermitian system, where all eigenstates are orthogonal to each other. In Fig.1 (a), the lattice is supposed to be gain/loss free and moderately nonreciprocal with site-independent hopping amplitudes, $\ds{t_n=2t_n^{\prime}=0.1}$. In this system with open edges, the non-Hermitian skin effect occurs and all eigenstates are localized around the left edge. Since the contrast between $\ds{t_n}$ and $\ds{t_n^{\prime}}$ is not large (the system is a bit away from the exceptional point), the densities of some eigenstates can be visually distinguished in the plot. As opposed to the case in (a), the spatial densities are tightly packed for (b,c), where we choose a different form of $\ds{t_n^{\prime}}$. Note that the corresponding densities for $12$ distinct eigenstates are so close to each other especially in (c) that one can hardly distinguish the densities for each eigenstates visually. The lattices in (b,c) are supposed to be closed to show that eigenstate clustering is not restricted to lattices with open edges. In (b), one can see that localization occurs at around $n=9$th lattice site. Since it is not the edge point, non-Hermitian skin effect can not be used here to explain this kind of dense localization of all eigenstates. We stress that no localization occurs in (c) as opposed to the other cases. Instead, all eigenstates are extended. So we say that clustering occurs also for extended states.
\begin{figure}[t]\label{2pxoik0}
\includegraphics[width=4.25cm]{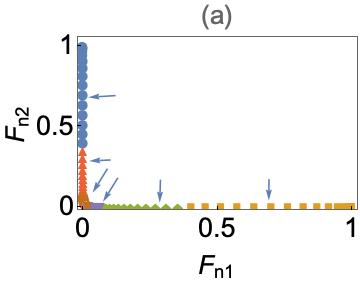}
\includegraphics[width=4.25cm]{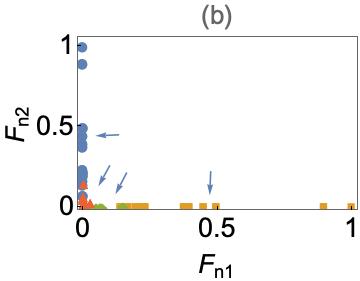}
\caption{ The plots  (a) and (b) show the classifications of eigenstates in k-groups according to k-means clustering algorithm for the same systems described in Fig 1 (a) and (b) respectively but with $N=80$. Different classes are represented in different colors. We take $k=6$ for (a) and $k=4$ for (b). The x-axis and y-axis are $\ds{   F_{n1}  }$ and  $\ds{    F_{n2}  }$, respectively.  }
\end{figure}\\
We can't predict eigenstate clustering just by looking at the densities of the eigenstates since even two orthogonal eigenstates can have the same density profiles. To study eigenstate clustering, we should calculate the corresponding fidelities to produce a set of data points and then use a clustering algorithm. Let us numerically calculate the fidelities for the system described in Fig. 1 (a,b) but with $N=80$. We use the lowest lying two eigenstates to form the data space: $\ds{ \{  | \alpha>\}= \{ \psi_1,\psi_2   \}   }$, where $\ds{F_{1,2}\approx 0}$ (they are almost orthogonal and hence dissimilar). Then we calculate fidelities with respect to these two reference eigenstates. In Fig. 2, we plot all the fidelities in the 2D data space and classify them in groups according to the standard k-means clustering algorithm. Different groups are represented in different colors in the plots. We take $k=6$ and $k=4$ for (a) and (b), respectively. In other words, the eigenstates in (a) and (b) are accumulated together in $k=6$ and $k=4$ classes because of certain similarities among them. The eigenstates in the same group are very similar to each other while the eigenstates in different group are different. For example, the eigenstates in the same group have almost the same form and their densities are hardly distinguishable. This implies that unavoidable disorder in an experiment can induce transition between them due to the slight difference in the form of the eigenstates in the same group. However, the real and imaginary parts of the eigenfunctions are quite different from each other in different groups even if their densities resemble to each other.\\
One can intuitively say that gain and loss lower the degree of the similarities between the eigenstates in a non reciprocal lattice, where the non-Hermitian skin effect occurs. Here we show that gain and loss can enhance this effect and  all the eigenstates can become more localized and more tightly packed. Consider a non-reciprocal lattice with balanced gain and loss $\ds{\gamma_n=(-1)^n \gamma}$ subject to the open boundary conditions. In Fig. 3, we plot the densities at a specific value of $\ds{\gamma}$ and the set of fidelities $\ds{F_{nm}}$ with $\ds{m{\neq}n}$ as a function of the non-Hermitian strength. Note that $\ds{F_{nm}=F_{mn}}$ and hence we show $\ds{N(N-1)/2}$ different combinations for a lattice with $N$ sites. The Fig. 3 (a) shows how the fidelities are distributed as a function of the non-Hermitian strength in a highly nonreciprocal lattice $\ds{t_n=10t_n^{\prime}=0.1}$. At $\gamma=0$ (a nonreciprocal lattice with no gain/loss), the fidelities are almost evenly distributed in between $0.25$ and $1$. Because of the high contrast between  $\ds{t_n }$ and  $\ds{ t_n^{\prime}}$ (the system is very close the exceptional point), there is no zero fidelity among eigenstates and hence the data space is 1D. In such a space, classification is trivial. It is interesting to see from Fig. 3 (a) that the set of the fidelities are contracted as $\gamma$ increases. This means that gain and loss can enhance the similarity between the eigenstates. At around $\gamma=0.04$, the eigenstates are maximally similar. This can also be seen from Fig. 3 (b) where all of the eigenstates are highly clustered around the left edge at the gain/loss strength $ \gamma=0.04$. As can be seen from Fig. 3 (a), a new brach appears at around $\gamma=0.04$ and some eigenstates are getting less and less similar to other ones as $\gamma$ is increased. The dimension of the corresponding data space becomes more than $1$ when $\ds{\gamma>0.3}$. 
\begin{figure}[t]\label{2678ik0}
\includegraphics[width=4.25cm]{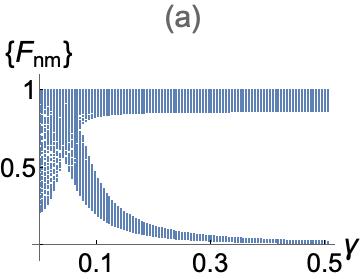}
\includegraphics[width=4.25cm]{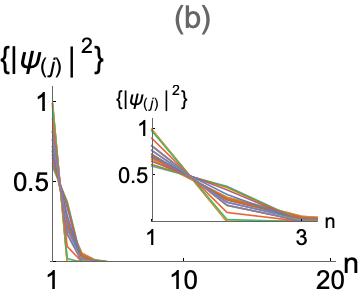}
\caption{ The set of all fidelities among the eigenstates with $\ds{n<m}$ as a function of $\ds{\gamma}$ ($\ds{\gamma_n=(-1)^n\gamma }$) for the nonreciprocal lattice with gain/loss (a): $\ds{  \{  F_{nm}  \} = \{  ~F_{12}, F_{13},...,F_{1N}, F_{23}, F_{24},...,F_{2N}, ...,F_{(N-1)N} ~ \}  }$ (There are $\ds{N(N-1)/2}$ elements in the set at a given value of $\ds{\gamma}$). The lattice has open edges and $\ds{t_n=0.1}$, $\ds{t^{\prime}_n=0.01}$ and $\ds{N=20}$. In the absence of gain/loss in this highly nonreciprocal lattice, all of the fidelities in the set are larger than $0.25$, which is the signal of non-Hermitian skin effect. As the gain/loss strength $\ds{\gamma}$ is increased up to $\ds{\gamma=0.04}$, the minimum fidelity in the set increases, which shows that gain/loss can enhance eigenstate similarity. However, for $\ds{\gamma>0.04}$, a branch decreasing with $\ds{\gamma}$ appears. This implies that some eigenstates become dissimilar to other ones for large values of gain/loss strength. In (b), we plot the densities for all eigenstates at $\ds{ \gamma= 0.04}$, where eigenstate similarities in the system are maximum. The $20$ eigenstates are so similar to each other that their densities can hardly be distinguished visually. In the inset, we also plot them up to $n=3th$ lattice sites for more clarity.}
\end{figure}\\
So far, we have studied classification of non-orthogonal eigenstates. Another question arises. Can we qualitatively discuss time evolution of an arbitrary initial wave packet $\ds{\Psi(t=0)}$? To study this problem, we consider the fidelities among the eigenstates as the training data. For example, the set of data in Fig. 2 can be training data. Next, we find the fidelity between $\ds{\Psi(0)}$ and $\ds{|\alpha>}$. These are our test data. The initial wave packet $\ds{\Psi(0)}$ is then placed in a specific group. Suppose $\ds{\Psi}$ sits in a particular class where all eigenstates in the same group have real eigenvalues. In this case, $\ds{\Psi}$ makes small amplitude power oscillation among the eigenstates in the same group. After a large time, the wave packet can make a transition to some other states in a different group. If, on the other hand, some of the eigenstates in the same group have complex eigenvalues, then even a small amplitude transition of $\ds{\Psi}$ into the state with highest imaginary part of the eigenvalue grows in time and becomes rapidly dominant. Note that there may exists an eigenstate in a different group whose imaginary part of eigenvalues is highest in the system. It takes some more time for the initial wave packet to become that eigenstate. To this end, we stress that our classification is not good enough to make such a prediction if $\ds{\Psi(0)}$ sits away from each clustering center in the data space.\\  
Let us briefly discuss topological features of clustered eigenstates. As opposed to Hermitian systems, topological eigenstates in non-Hermitian systems are not orthogonal to bulk eigenstates. Depending on a specific non-Hermitian system, topological eigenstates can be clustered in the same group as some bulk states. This implies that weak disorder can induce transition among the eigenstates clustered in the same group. This leads to the breakdown of chirality of topological states in 2D as discussed by our earlier paper \cite{tevolv}. This is because of the fact that propagation of topological state is supported in only one direction in an Hermitian 2D strip, since no state is available at the same energy that propagates in the opposite direction on the same edge. But this is not the case as there are many other eigenstates (localized on the same edge and with close energy eigenvalues) in the same class. Therefore, the topological eigenstates are no longer robust in this system. As a result, we say that topological features can be broken in non-Hermitian systems if eigenstate clustering occurs.\\
To sum up, we have predicted eigenstates clustering and systematically studied clustering of non-orthogonal eigenstates. We have discussed that non-Hermitian skin effect is a typical example of eigenstate clustering. To the best of our knowledge, our paper is the first paper in the literature to use a clustering algorithm in non-Hermitian systems. We have used fidelities and perform k-means clustering algorithm already used in machine learning to classify clustered eigenstates in groups according to their similarities. We think our paper will pave the way for the usage of clustering and classification algorithms and machine learning in non-Hermitian systems.

\end{document}